\documentclass[preprint2]{emulateapj}
\usepackage{color}
\usepackage{amsmath}
\slugcomment{published in The Astrophysical Journal 729 (2011) 91, Received 2010 November 25; accepted 2011 January 5}
\shorttitle{Photoproducts of Diamondoids}
\shortauthors{Steglich et al.}

\begin{document}
\title{Electronic spectroscopy of FUV-irradiated diamondoids:\\
A combined experimental and theoretical study}
\author{M. Steglich\altaffilmark{1} \& F. Huisken}
\affil{Laboratory Astrophysics Group of the Max Planck Institute for Astronomy at the Friedrich Schiller
University Jena,\\
Institute of Solid State Physics, Helmholtzweg 3, D-07743 Jena, Germany}
\email{M.Steglich@web.de}
\author{J. E. Dahl}
\affil{Geballe Lab for Advanced Materials, Stanford University, 476 Lomita Mall, Stanford, CA 94305, USA}
\author{R. M. K. Carlson}
\affil{MolecularDiamond Technologies, Chevron Technology Ventures, 100 Chevron Way, Richmond, CA 94802, USA}
\author{Th. Henning}
\affil{Max Planck Institute for Astronomy, K\"onigstuhl 17, D-69117 Heidelberg, Germany}
\altaffiltext{1}{Corresponding author}

\begin{abstract}
Irradiation with high energy photons (10.2 $-$ 11.8 eV) was applied to small diamondoids isolated in solid rare gas matrices at low temperature. The photoproducts were traced via UV absorption spectroscopy. We found that upon ionization the smallest of these species lose a peripheral H atom to form a stable closed-shell cation. This process is also likely to occur under astrophysical conditions for gas phase diamondoids and it opens the possibility to detect diamond-like molecules using their rotational spectrum since the dehydrogenated cations possess strong permanent dipole moments. The lowest-energy electronic features of these species in the UV were found to be rather broad, shifting to longer wavelengths with increasing molecular size. Calculations using time-dependent density functional theory support our experimental findings and extend the absorption curves further into the vacuum ultraviolet. The complete $\sigma - \sigma^*$ spectrum displays surprisingly strong similarities to meteoritic nanodiamonds containing 50 times more C atoms.
\end{abstract}


\section{INTRODUCTION}
Diamond-like material is expected to be abundant in the interstellar medium \citep{Henning98}. Small nanodiamonds (2$-$3 nm) were extracted from meteoritic material, and they are the most abundant presolar grains in the primitive meteorites \citep{lewis87, anders93, jones04}. Recently, diamondoid molecules were the subject of different experimental and theoretical studies revealing their spectroscopic properties \citep{oomens06, bauschlicher07, lenzke07, pirali07, landt09a, landt09b}. These special hydrocarbons can be considered as faced-fused diamond cages where all carbon atoms are sp$^3$ hybridized with hydrogens saturating the dangling bonds on the surface. The diamond-like structure results in a remarkable rigidity, strength, and thermodynamic stability, especially compared to other hydrocarbons. The smallest of these species, adamantane C$_{10}$H$_{16}$, consists of only one diamond cage, followed by diamantane C$_{14}$H$_{20}$ with two, and triamantane C$_{18}$H$_{24}$ with three faced-fused cages. Diamondoids can be found in some natural gas reservoirs, and especially diamantane is one of the deposits in gas pipelines \citep{reiser96}. Lately, diamondoids consisting of up to 11 diamond cages were isolated from petroleum by \citet{dahl03} making these species accessible to laboratory investigations.

\citet{oomens06} measured the infrared spectroscopic properties of powders of higher diamondoids (up to hexamantane) at room temperature. In accordance with calculations applying density functional theory (DFT), the by far strongest features in the IR spectra of neutral diamondoids were found to be the C-H stretching bands between 3.4 and 3.6 $\mu$m arising from the hydrogen-terminated surfaces. \citet{pirali07} additionally measured the IR emission spectra of hot (500 K) gas phase adamantane, diamantane, and triamantane in the wavelength region of the C-H stretching modes revealing a small redshift of the bands in the solid-state spectra. Based on the IR spectra obtained by \citet{oomens06} along with additional DFT calculations, \citet{pirali07} also made assignments for two IR features of two different classes of astronomical objects. The first one is the unusual IR emission at 3.43 and 3.53 $\mu$m originating from two objects whose spectra are elsewhere dominated by the well-known infrared emission bands of polycyclic aromatic hydrocarbons (PAHs), Elias 1 and the inner region of the circumstellar disk around HD 97048 \citep{Habart04}. Although these bands were already assigned to nanodiamonds of at least 50 nm diameter \citep{guillois99}, it was shown that also tetrahedral diamondoid molecules containing around 130 C atoms (close to the size of the smallest meteoritic nanodiamonds) exhibit the 3.43 and 3.53 $\mu$m bands with the proper intensity ratio \citep{pirali07}. The second feature mentioned is the broad (FWHM 0.09 $\mu$m) absorption band centered at 3.47 $\mu$m which is observed in the absorption spectra of various dense clouds in lines of sight toward young stellar objects \citep{allamandola92, allamandola93}. By co-adding all diamondoid solid-state spectra obtained by \citet{oomens06}, \citet{pirali07} showed that the different C-H stretching modes merge into a broad structure centered around 3.47 $\mu$m, and they argued that small diamond-like molecules may therefore be a major contributor to the interstellar absorption band which is only observed in or behind dense molecular clouds, but not in the diffuse interstellar medium. As the intensity of the interstellar band could be correlated with the intensity of the 3.08 $\mu$m band of water ice, it was speculated that its carriers may be formed on icy grains in the shielded environment of molecular clouds \citep{brooke96} which is furthermore supported by recent laboratory experiments where nanodiamond crystallites were created upon UV irradiation of interstellar ice analogs \citep{kouchi05}. Based on the computed intensities of the diamondoid C-H stretching bands, \citet{bauschlicher07} deduced that only 1$-$3\% of the cosmic C has to be locked in diamondoids in order to explain the observed intensity of the interstellar band. Furthermore, they calculated ionization potentials (IPs) of neutral diamondoids, as well as IR spectra and electronic transition energies of neutral and cationic diamondoids, concluding that cations may also contribute to the 3.47 $\mu$m absorption band. However, the bands of the cations are somewhat weaker than those of their neutral counterparts. Furthermore, the cations feature further bands with comparable strengths at longer wavelengths (6$-$18 $\mu$m), e.g. due to C-H bending vibrations, which could be used in the future to trace ionized diamondoids.

Compared to the strong IR emission of PAHs triggered by the absorption of UV-vis photons, the IR emission of neutral diamondoids is rather inefficient because their electronic absorption onset lies far in the UV between 6 and 7 eV \citep{landt09a, landt09b}. Calculations imply that diamondoid cations would feature absorption bands in the visible and near-IR due to their open-shell structure \citep{bauschlicher07}. However, these transitions are very weak and, as for the neutrals, efficient IR emission can only be expected in regions of space experiencing high-energy and high-flux UV radiation fields, usually in close proximity of the exciting stars \citep{bauschlicher07}. Therefore the 3.43 and 3.53 $\mu$m emission features are so rarely observed, to date only in HD 97048 and Elias 1.

The IP of diamondoids \citep[8$-$9 eV;][]{lenzke07} is only slightly higher than their band gap. \citet{lenzke07} have shown that the photoion yield of diamondoids reaches its maximum between 10 and 11 eV, almost exactly around the hydrogen Ly$\alpha$ emission. Hence, ionization of neutral diamondoids in strongly irradiated regions of space (HD 97048 and Elias 1) may be very efficient. Consequently, one should address the following questions: are these cationic species stable and, if so, what are their spectroscopic fingerprints? As already stated, one can expect weak absorption bands in the visible and near-UV wavelength range for the cations due to low-energy transitions to semi-occupied molecular orbitals, which might be detectable by electronic spectroscopy methods. Unlike in PAHs, there is no delocalized electron cloud. Removing one electron due to photoionization weakens the bonds between the atoms, and fragmentation (C$-$H bond breaking) may occur. Indeed, such behavior was observed for the three smallest diamondoids. By comparing the IR absorption bands of positively charged, gas phase adamantane \citep{polfer04}, diamantane, and triamantane \citep{pirali10} with theoretical spectra of dehydrogenated diamondoid cations, it was shown that these molecules easily lose a hydrogen atom upon ionization to form stable closed-shell species. The loss preferentially happens on a tertiary carbon (CH group) rather than on a secondary carbon (CH$_2$ group). Since in these experiments, the ionization was performed via charge transfer using cationic agents with high IPs, it is not obvious whether this dehydrogenation will also occur under astrophysical irradiation conditions.

In this work, we investigated the electronic transitions of the four smallest diamondoids, namely adamantane C$_{10}$H$_{16}$, diamantane C$_{14}$H$_{20}$, triamantane C$_{18}$H$_{24}$, and tetramantane C$_{22}$H$_{28}$ (three isomers), and their photoproducts. For this purpose, we used matrix isolation spectroscopy (MIS). Cationic species were formed via UV irradiation using a hydrogen-flow discharge lamp to simulate the interstellar UV photon field. Our experimental findings are supported by theoretical calculations applying DFT and time-dependent DFT (TD-DFT).  These results will help astronomers to search for the spectroscopic fingerprints of diamond-like molecular species. Furthermore, the UV absorption cross sections of neutral and cationic diamondoids can be used to accurately predict IR emission processes caused by stochastic heating due to the absorption of UV photons in strongly irradiated regions of space.

\section{METHODS}
\label{methods}

\subsection{Theoretical Calculations of Electronic Spectra} \label{theory}
For the purpose of identification and comparison, we performed DFT calculations for differently charged and de-hydrogenated diamondoids. The molecular structures were first optimized with the Gaussian09 software \citep{frisch09} using the B3LYP functional \citep{stephens94, becke93} in conjunction with the 6-311$++$G(2d,p) basis set for adamantane and its derivatives, and the 6-311$+$G(d) basis set for the larger diamondoids, respectively. We chose the smaller basis set for the larger molecules to reduce the computational effort since no remarkable differences between 6-311$++$G(2d,p) and 6-311$+$G(d) were obtained for the ground-state structure and electronic spectrum of adamantane. The vibrational modes were calculated afterward to check whether the optimized structures were really at their respective minima of the potential energy surfaces and to determine the zero-point corrections for the ground-state energies. The dipole moments presented later have been corrected to correspond to the center-of-mass coordinate system. The values obtained by the Gaussian09 software refer to molecular orientations with the center of nuclear charge as origin. For the investigated species, both values differ by less than 5\%. Two different approaches were used to evaluate the energies of the excited electronic states. The first one involves the TD-DFT implementation of Gaussian09, using the same functional and basis set as for the ground-state calculations. The computational effort scales steeply with the size of the system under consideration and the number of excited states to be calculated. This TD-DFT approach was used to accurately predict the first few electronic transitions of the diamondoids and their derivatives. An error of about 0.3 eV regarding the energetic positions of the transitions can commonly be expected at this level of theory. It should be mentioned that purely vertical electronic transitions are calculated. Vibrational excitations in excited electronic states cannot be accounted for. However, for higher energies the lifetimes of the excited states are expected to decrease substantially resulting in broad bands without discernible vibrational pattern even for cold gas phase molecules.

For the high-energy states, the TD-DFT formalism as implemented in the Octopus software package \citep{castro06, marques04} was applied. Unlike the frequency-space implementation of TD-DFT in Gaussian09, Octopus uses real-space numerical grids to propagate the Kohn Sham orbitals in real time. Followed by an initial very short electric pulse, exciting all frequencies of the system, the time-dependent dipole moment is calculated from which the linear optical absorption spectrum can be derived. Relying on numerical meshes, the code works without basis sets. However, we used again the B3LYP functional. The volume of the box in which the desired molecule is represented was chosen such that each atom was at least 4 \AA\ away from the edges. In all calculations, the grid spacing was 0.2 \AA, the time integration length was 10 $\hbar$ eV$^{-1}$, and the time step was 0.002 $\hbar$ eV$^{-1}$. This approach does not yield any information on the symmetry of the excited states. The widths of the absorption bands are purely artificial and depend solely on the integration length used in the calculation. However, the area of each band is directly related to the oscillator strength of the corresponding transition and can easily be converted into absorption cross section values. Optical spectra derived from the Octopus code already gave reasonable agreement between calculated and measured absorption spectra and cross sections of PAHs in the energy range above the IP (10$-$30 eV) where electronic transitions involving $\sigma$ electrons dominate \citep{malloci04}. It should be stressed that only bound-bound transitions are predicted by this approach, whereas transitions leading to direct ionization cannot be accounted for. However, in view of the good agreement between experiment and theory for PAHs, it was argued that superexcited states account for most of the absorption in the vacuum ultraviolet (VUV) and that they are coupled to the ionization continuum \citep{malloci04}. Whether the same situation also applies for the $\sigma - \sigma^*$ transitions of diamondoids and their ionic derivatives can only be clarified when dedicated laboratory measurements are available.

\subsection{Matrix Isolation Spectroscopy and FUV irradiation} \label{misfuv}
\begin{figure}[t]\begin{center}
\epsscale{1.1} \plotone{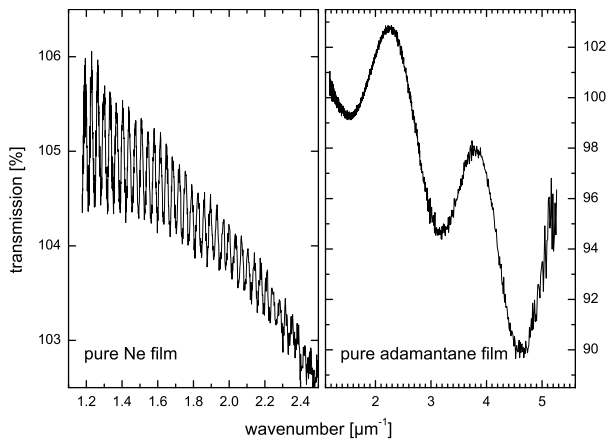} \caption{Transmission spectra of pure Ne (left) and adamantane (right) on a 6.8 K cold CaF$_2$ window used for the determination of the deposition rates and hence the sample to matrix ratio. The thickness $d$ of the deposited film follows from the wavenumber difference $\Delta k$ between two consecutive transmission maxima via $\Delta k = (2nd)^{-1}$, where $n$ is the film's refractive index. Note the increased scattering toward shorter wavelengths. The transmission is actually higher than 100\% because it is compared to the transmission of the clean CaF$_2$ window which has a higher reflectance than the deposited films.} \label{fig1}
\end{center}\end{figure}

To measure low-temperature spectra of the diamondoids and their photoproducts, we used a setup for MIS. A transparent CaF$_2$ window mounted inside of a vacuum chamber at the bottom of an expander was cooled down to temperatures $<$7 K. The molecules to be examined were condensed onto it together with an excess of inert gas. As matrix material we used neon (Linde, purity 99.995 \%) whose polarizability is low enough to keep the perturbation of the investigated molecules to a minimum. By turning the CaF$_2$ window by 90$^\circ$, transmission spectroscopy down to 190 nm was performed with a spectrophotometer (JASCO V-670 EX), calibrated with an accuracy of 0.3 nm. The chosen resolution was typically 0.5 nm, which is much smaller than the widths of the measured absorption bands. Prior to the incorporation of the diamondoids into the matrix, they were evaporated in an oven kept at suitable temperatures. Due to the high vapor pressures of the smallest diamondoids, the oven had to be cooled down to temperatures as low as 0$^\circ$C in the case of adamantane and diamantane. For adamantane, a special polyvinylidene fluoride filter further decreased the gas flow to the CaF$_2$ window. Triamantane was heated up to 20$^\circ$C $-$ 30$^\circ$C and the tetramantanes to 90$^\circ$C. The deposition rates of the neutral diamondoid precursors and the matrix material, both transparent in the investigated wavelength range, were determined in separate experiments by deposition of the pure substance on the low-temperature window and measuring the transmission of the as-prepared films with the spectrometer. A typical interference pattern corresponding to the transmission of an etalon could be observed for sufficiently thin films from which the thicknesses and, therefore, the column densities could be derived (see Figure \ref{fig1}).

After preparation of the transparent matrix doped with the neutral precursor molecules, an absorption spectrum, usually displaying no discrete features, was measured and used as a new baseline. The samples were then photolyzed with the far-UV emission from a microwave-driven hydrogen-flow discharge lamp \citep{warneck62}. The spectra recorded afterward\footnote{All experimental spectra presented here were obtained by using the spectra recorded before irradiation as a baseline.} feature absorption bands from species created during the irradiation. For comparison, we also irradiated a clean Ne matrix and films of pure diamondoids which were deposited on the 7 K cold CaF$_2$ window without simultaneous Ne flow. Besides trace amounts of dissociated water in the case of the inert gas, we did not observe any features. This confirms that the measured bands are indeed due to photoproducts of isolated diamondoids. The hydrogen lamp itself was operated with a gas mixture of 10\% H$_2$ and 90\% Ar, usually at a pressure of 0.6 mbar. The purpose of the inert gas is to suppress the 160 nm molecular emission and enhance the 121.6 nm (10.2 eV) Ly$\alpha$ emission, but it also introduces additional lines from the Ar at 106.7 nm (11.6 eV) and 104.8 nm (11.8 eV). The lamp was separated from the vacuum of the MIS chamber by a LiF window. Its total photon flux of $10^{13}-10^{14}$ photons s$^{-1}$ was determined by measuring the photocurrent emanating from a Pt plate under FUV light exposure. During the experiments, the intensity on the sample surface was typically $10^{15}-10^{16}$ photons m$^{-2}$ s$^{-1}$. Usually, the irradiation was performed for 15$-$30 minutes.

\section{RESULTS AND DISCUSSION}
\label{results}

\subsection{Adamantane} \label{adam_section}
\begin{figure}[t]\begin{center}
\epsscale{1.1} \plotone{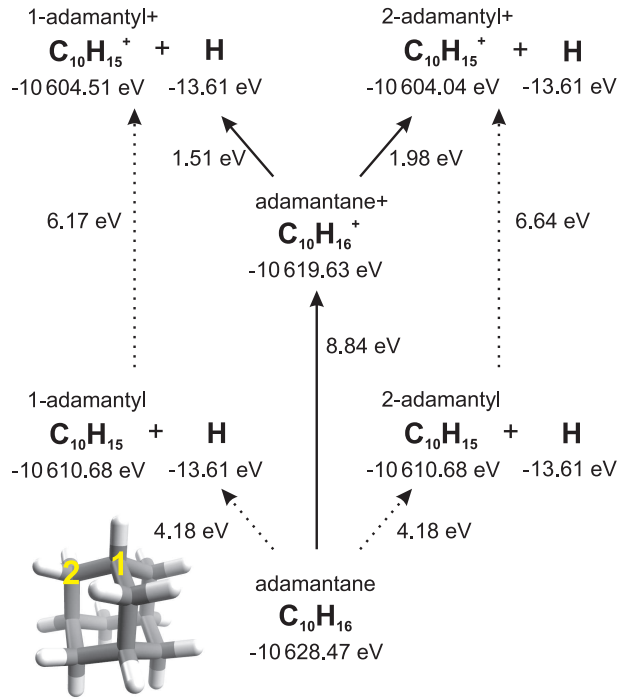} \caption{Zero-point-corrected ground-state energies of adamantane and its derivatives calculated at the B3LYP/6$-$311$++$G(2d,p) level of theory. The formation routes of the adamantyl cations in the experiment are indicated by solid arrows.} \label{fig2}
\end{center}\end{figure}

In Figure \ref{fig2}, the structure and calculated ground-state energy of adamantane, as well as the energies of the adamantane and adamantyl cations are displayed. Adamantane possesses two structurally different H sites and therefore two different adamantyl isomers. The numbers on the adamantane structure shown in Figure \ref{fig2} indicate the C atoms from which the H atoms are removed to form the corresponding adamantyl structures. According to the calculations, only minor distortions of the carbon framework are expected upon H removal. This also applies to the larger species we have investigated. For the ground states of the adamantyl cations, different spin states are principally possible. However, the triplet state of the 1-adamantyl cation is about 3 eV higher in energy than the singlet state. Because of this rather high energy difference, we expect all other dehydrogenated diamondoid cations to be closed-shell singlet species as well and the calculations were performed accordingly. For completeness, we also calculated the structures and ground-state energies of the neutral 1- and 2-adamantyl radicals. In both cases, an energy of 4.2 eV would be necessary to remove one H atom. However, direct photodissociation is unlikely since the absorption onset is further in the UV. Basically, the photons delivered by the hydrogen lamp provide enough energy to cause photoionization, as well as the removal of one H atom from the ionized molecule. Further dissociation cannot be accomplished with a single photon. Considering the low FUV doses applied during the experiments, comprising typically 15$-$30 minutes of irradiation, further processing of already ionized molecules can be excluded. This was ensured by increasing the irradiation time up to 2 hr without noticeable change of the absorption bands.

\begin{figure}[t]\begin{center}
\epsscale{1.15} \plotone{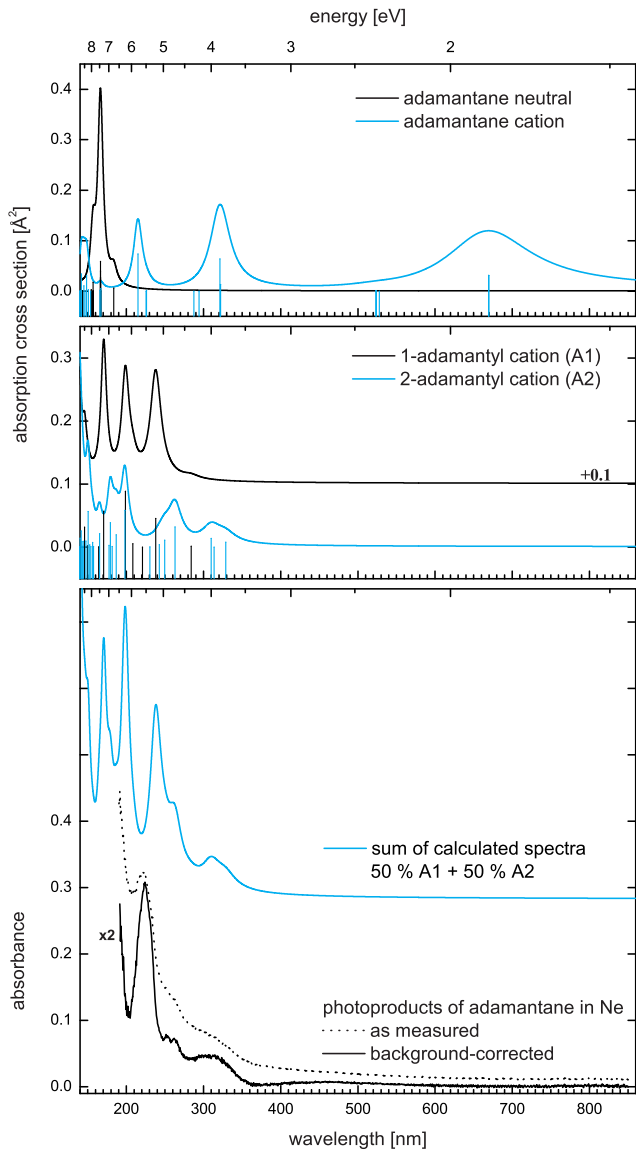} \caption{Calculated (B3LYP/6$-$311$++$G(2d,p)) spectra of neutral and cationic adamantane (top panel) and singly dehydrogenated cationic adamantyl (middle panel). The measured spectrum of FUV-irradiated adamantane isolated in solid Ne (6.8 K) is displayed in the bottom panel. Note that the calculated spectra are just representations of the calculated excited states (stick spectra in top and middle panels). They have been computed by convolution using Lorentzians with a width of 3000 cm$^{-1}$ to indicate the range where strong electronic transitions can be expected.} \label{fig3}
\end{center}\end{figure}

The calculated spectra of neutral and cationic adamantane and adamantyl, using the B3LYP/6$-$311$++$G(2d,p) level of theory, are displayed in the upper two panels of Figure \ref{fig3}. The theoretical spectra were obtained by convolving the shown stick spectra, representing the oscillator strength of each transition at its corresponding transition wavelength, with Lorentzian functions. While the area of each Lorentzian is proportional to the calculated oscillator strength, the width was chosen to be constant (3000 cm$^{-1}$). The chosen bandwidth is rather arbitrary, but the so-computed spectrum indicates in which energy range strong electronic transitions of the real molecule can be expected. 

The measured spectrum of FUV-processed adamantane in solid Ne (6.8 K) can be found in the bottom panel of Figure \ref{fig3}. The ratio of Ne to adamantane in terms of absolute numbers of atoms or molecules was determined as described in Section \ref{misfuv}. The densities of solid Ne and adamantane were taken to be 45 atoms nm$^{-3}$ \citep{timms96} and 1.2 g cm$^{-3}$ \citep{yashonath86}, respectively. Using these values, the isolation ratio (Ne to adamantane) was estimated to be better than 190 $n_{\text{Ada}}$ $n^{-1}_{\text{Ne}}$. The factor $n_{\text{Ada}}$ $n^{-1}_{\text{Ne}}$ is the ratio of the refractive indices of the pure solid materials in the visible and should be between 1 and 2. Nevertheless, we also performed measurements at lower isolation ratios and did not notice appreciable spectral differences. Since it is unknown what fraction of the neutral precursor molecules is actually transformed into cations\footnote{Usually, the conversion rate is lower than 10\% under these experimental conditions.} we cannot provide experimental values for the absorption cross section.

As indicated in Figure \ref{fig3}, an additional baseline correction was applied to remove the strong scattering background in the UV. Because of its wavelength dependence ($\sim \lambda^{-4}$), we attribute this background mainly to an increased Rayleigh scattering of the charged species compared to their neutral counterparts. The derived absorption spectrum consists of four broad features above 200 nm. The broad band between 280 and 350 nm with maximum around 308 nm is not an artifact of the measurement. It has been confirmed by repeatedly performed experiments. The same applies to the two somewhat narrower features at 252 and 261 nm. The strongest band in the accessible wavelength region extends from 200 to 240 nm and peaks at 223.5 nm. Considering the theoretical results, we exclude the presence of open-shell adamantane cations in the photo-processed matrix due to the absence of bands in the visible. Instead, the observed spectrum points toward the creation of the more stable, closed-shell, singly de-hydrogenated cation. Therefore, we assign the strongest band at 223.5 nm to the S$_0 \rightarrow$ S$_2$ transition of the 1-adamantyl cation (point group C$_{3\text{v}}$) which is the isomer with lower ground-state energy. The calculated oscillator strength of this band is $f = 0.091$. Its position is 0.3 eV away from the measured value in Ne matrix which is a reasonable error for the applied theoretical model. The other weaker transitions may be related to the S$_0 \rightarrow$ S$_1$ transition of the 1-adamantyl cation ($f = 0.0028$) and, more likely, to the first four excited states of the 2-adamantyl cation (point group C$_{\text{s}}$) with calculated oscillator strengths below $f = 0.033$. A more reliable assignment on the basis of computed spectra would be possible via IR spectroscopic investigations, because IR-active vibrations can be more easily and accurately calculated using quantum chemical models. However, this is beyond the scope of the present investigation. Nevertheless, taking into account the weaker band strengths of the 2-adamantyl cation and comparing the calculated with the measured spectra, it seems that the second isomer is almost as abundantly created as the 1-adamantyl cation. The bottom panel of Figure \ref{fig3} contains a sum spectrum of both isomers (each contributing 50 \%). Besides the 0.3 eV redshift of the 1-adamantyl band, it closely resembles the measured spectrum. To some extent, this would be in contradiction to the results obtained by \citet{polfer04} who used an indirect charge transfer method to create the ions and subsequently observed only the isomer with the lowest ground-state energy. A possible explanation may be found in the different experimental techniques (charge transfer versus photoionization) and conditions (gas phase versus matrix at low temperature) that were applied.

Considering the previous results, the adamantane molecules were subjected to dissociative photoionization upon irradiation with FUV photons of energy 10.2 $-$ 11.8 eV according to
\begin{equation}
\text{C}_{10}\text{H}_{16} + h \nu \rightarrow \text{C}_{10}\text{H}_{15}^{+} + \text{e}^{-} + \text{H}.
\end{equation}
This process has been described for a few smaller molecules, e.g., H$_2$O \citep{cairns71} or CH$_4$ \citep{samson89}. After ionization, some excess energy is stored in the vibrational degrees of freedom of the cationic molecule which subsequently leads to the destruction of one terminal C$-$H bond. During this process, electrons and neutral H atoms are released which are usually trapped on defects or impurities in the matrix. Because of recombination reactions between positively charged molecules and released electrons (and H atoms), the ion yield saturates when a certain irradiation dose is reached. The formation of negatively charged adamantane or neutral adamantyl molecules due to electron attachment can be ruled out for the following reasons. First, these species possess an open-shell electronic structure and, like the adamantane cation, would feature absorption bands in the visible part of the spectrum. And second, as far as neutral adamantane is concerned, the negative electron affinity \citep{drummond07} hampers further electron attachment.

Notably, we observed neither sharp absorption bands nor a clear vibrational pattern. We want to remark that the measured bands are much broader than what is expected for typical matrix-induced broadening (at least for 7 K neon matrices) which is mainly due to site effects.\footnote{Molecules in different sites of the matrix exhibit different redshifts of their absorption bands, effectively resulting in a broadening.} We tentatively attribute this to an intrinsic property of the molecule, i.e., a very short lifetime of the excited state which is not entirely caused by the interaction with the rare gas atoms. This would lead to the conclusion that the main difference to astrophysically more relevant spectra of cold gas phase adamantyl cations is a small matrix-induced redshift, but not a broadening of the absorption bands. By coincidence, the spectral shape is in surprisingly good agreement with the computed spectra of the purely electronic (vertical) transitions. Therefore, the absolute values of the absorption cross sections, as they appear in the calculated spectra, may be regarded as representative for real gas phase molecules.\footnote{Otherwise, only the integrated cross section or the oscillator strength could be taken.}

Finally, we want to highlight another important property of the adamantyl cations. Unlike their neutral precursor adamantane, these species possess rather strong permanent dipole moments pointing from the molecular center toward the C atom from which the H atom has been removed. The calculated dipole moments of the 1- and 2-adamantyl cations amount to 0.96 and 2.57 Debye, even stronger than the dipole moment of the open-shell cation (0.61 Debye). This opens the possibility to detect and identify molecular diamond-like species in space using their rotational spectra.

\subsection{Diamantane}

\begin{figure}[t]\begin{center}
\epsscale{1.0} \plotone{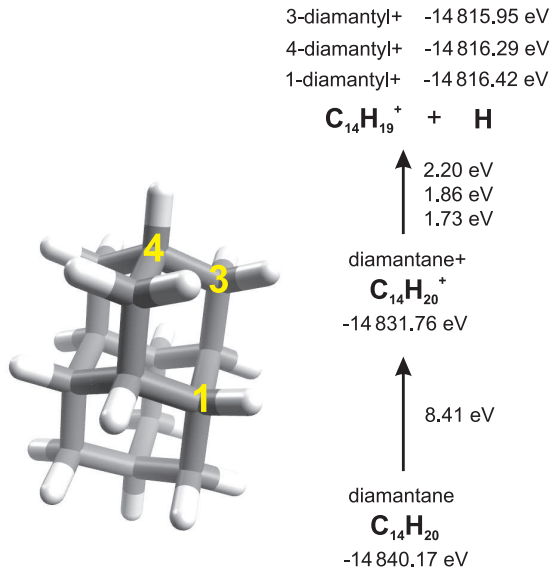} \caption{Zero-point-corrected ground-state energies of diamantane and its derivatives calculated at the B3LYP/6$-$311$+$G(d) level of theory. The diamantyl structures are labeled according to the IUPAC numbering system for diamondoids.} \label{fig4}
\end{center}\end{figure}

The calculated ground-state energies of diamantane and its singly dehydrogenated cationic derivatives are displayed in Figure \ref{fig4}. Again, the numbers on the diamantane structure shown indicate the positions from where the hydrogen atoms are removed to form the corresponding closed-shell cations. There are three possible isomers for the diamantyl cation, possessing quite strong permanent dipole moments of 1.76 (D1), 3.09 (D4), and 3.71 Debye (D3) due to the localized charge at the edge of the molecule.

\begin{figure}[t]\begin{center}
\epsscale{1.15} \plotone{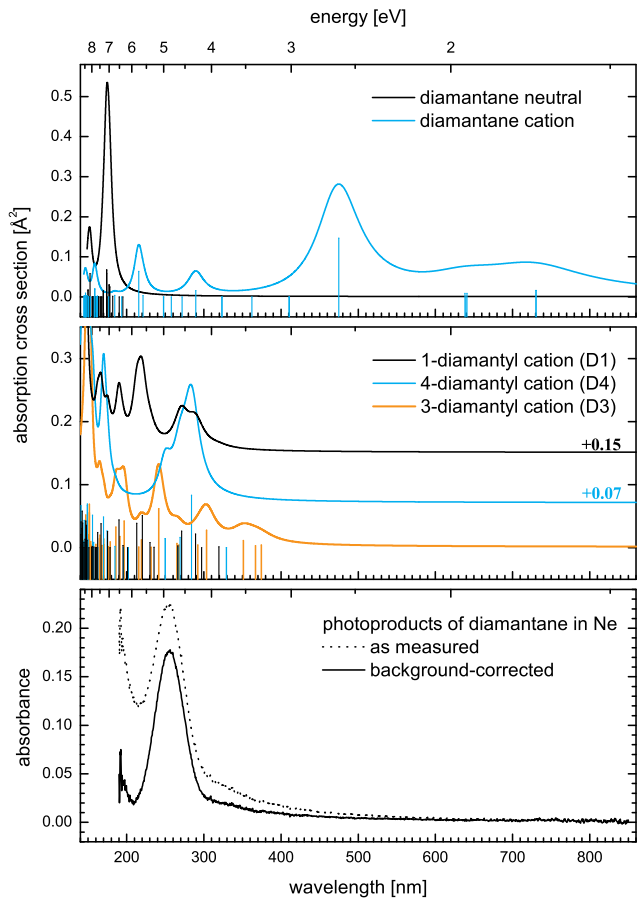} \caption{Calculated (B3LYP/6$-$311$+$G(d)) spectra of neutral and cationic diamantane (top panel) and singly dehydrogenated cationic diamantyl (middle panel). The measured spectrum of FUV-irradiated diamantane isolated in solid Ne (6.8 K) is displayed in the bottom panel.} \label{fig5}
\end{center}\end{figure}

The electronic spectra of diamantane and its related species are presented in Figure \ref{fig5}. An additional background correction has been applied on the red side of the measured spectrum to remove non-reproducible bumps and fringes due to baseline variations. Using 1.2 g cm$^{-3}$ \citep{karle65} as mass density of the solid diamantane deposit, the isolation ratio (Ne to diamantane) varied in different experiments between 450 $n_{\text{Dia}}$ $n^{-1}_{\text{Ne}}$ and 750 $n_{\text{Dia}}$ $n^{-1}_{\text{Ne}}$, where $n_{\text{Dia}}$ is the refractive index of the diamantane film. Compared to the photoprocessed adamantane, the spectrum of irradiated diamantane reveals a slightly broader peak (7800 cm$^{-1}$ versus 5900 cm$^{-1}$), positioned further to the red at 255 nm. As is evident upon inspection of the calculated and measured spectra, this feature cannot be explained by the presence of open-shell cation radicals. Regarding the formation of negatively charged diamantane or neutral diamantyl radicals, the same reasoning applies as in Section \ref{adam_section}. Principally, the photons from the hydrogen lamp carry enough energy to induce dissociative photoionization and create all three diamantyl isomers. However, the conclusion may be drawn that the main photoproduct in the matrix experiment is the 4-diamantyl cation (point group C$_{\text{3v}}$). Its first strong transitions S$_0 \rightarrow$ S$_{2,3}$ are predicted at 284 nm ($f=0.084$) and 269 nm ($f=0.033$), 0.5 and 0.26 eV away from the peak maximum of the measured band at 255 nm. Alternatively, the 255 nm band may partly or completely originate from transitions caused by the 1-diamantyl cation (C$_\text{s}$) for which several close-lying absorptions at 289 nm (S$_0 \rightarrow$ S$_3$, $f=0.022$), 271 nm(S$_0 \rightarrow$ S$_4$, $f=0.027$), and 266 nm (S$_0 \rightarrow$ S$_5$, $f=0.003$) are predicted. Due to the lack of certain absorption features in the measured spectrum, the presence of the 3-diamantyl isomer (C$_1$) in the matrix, i.e. the H removal from a CH$_2$ group, can rather be excluded. Because of the apparent absence of fine structure, it seems difficult to draw further conclusions.

\subsection{Triamantane}
\begin{figure}[t]\begin{center}
\epsscale{1.0} \plotone{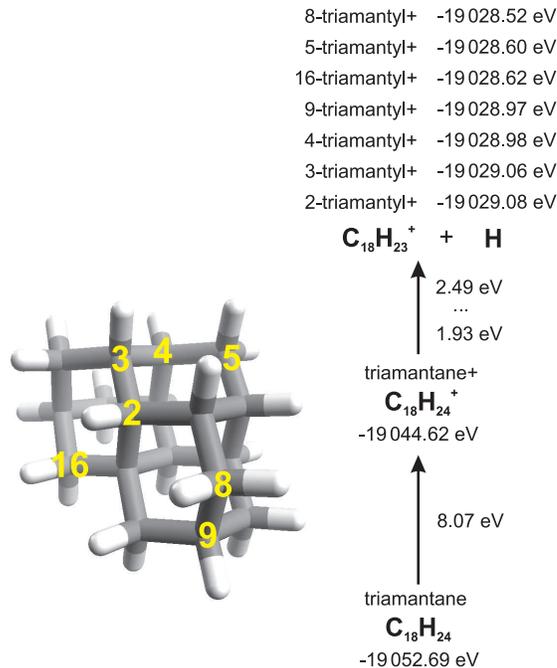} \caption{Zero-point-corrected ground-state energies of triamantane and its derivatives calculated at the B3LYP/6$-$311$+$G(d) level of theory. The triamantyl structures are labeled according to the IUPAC numbering system for diamondoids.} \label{fig6}
\end{center}\end{figure}

Figure \ref{fig6} displays the calculated ground-state energies of triamantane and its seven isomers of singly dehydrogenated cations. The necessary energies to remove one electron and one H atom from the parent molecule are comparable to the previously discussed diamondoids. Basically, the FUV lamp delivers photons with energies high enough to create all seven isomers. Their dipole moments, again quite strong, range between 1.13 and 5.72 Debye. The dipole moment of the open-shell cation amounts to 0.51 Debye.

\begin{figure}[t]\begin{center}
\epsscale{1.15} \plotone{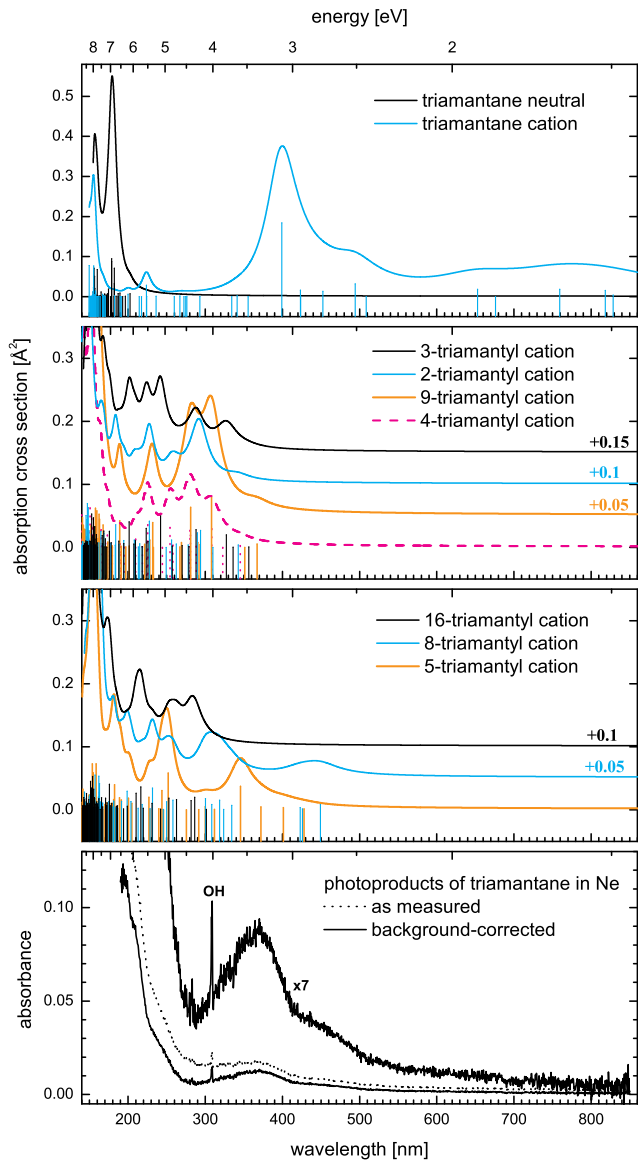} \caption{Calculated (B3LYP/6$-$311$+$G(d)) spectra of neutral and cationic triamantane (top panel) and singly dehydrogenated cationic triamantyl (two middle panels). The measured spectrum of FUV-irradiated triamantane isolated in solid Ne (6.8 K) is displayed in the bottom panel.} \label{fig7}
\end{center}\end{figure}

The corresponding calculated electronic spectra, as well as the measured spectrum of FUV-irradiated triamantane in Ne are displayed in Figure \ref{fig7}. The isolation ratio in the matrix experiment was on the order of 500 $-$ 1000 $n_{\text{Tria}}$ $n^{-1}_{\text{Ne}}$ with unknown refractive index of the pure triamantane ($n_{\text{Tria}}$) film. The photodissociation of trace amounts of water in the matrix is responsible for the narrow bands from the OH radical at 308 and 283 nm \citep{tinti68}. Compared to the previous measurements of irradiated adamantane and diamantane, the lowest-energy spectral feature shifts further to the red, extending roughly from 300 to 500 nm. It peaks at 368 nm and has a shoulder around 450 nm. A strong FUV rise also slides into the accessible wavelength region. An assignment to a certain isomer of the triamantyl cation is rather difficult. Oddly, the best match seems to be possible with the calculated spectrum of the 5-triamantyl cation which is 10.56 eV higher in energy than the triamantane neutral. (In the previously discussed measurements, the strongest bands seemed to be caused by species which were 10.35 eV (adamantane) and 10.27 eV (diamantane) away from the parent molecule.) Besides dehydrogenated triamantyl cations, an alternative explanation for the measured broad band would be the formation of the open-shell triamantane cation as can be seen from the comparison with the corresponding calculated spectrum. Possibly, triamantane is already big enough, and the energy stored in the molecule upon absorption of an FUV photon is distributed over sufficient vibrational modes to avoid H abstraction. Nevertheless, a clear identification of the created species on the basis of electronic absorption spectroscopy is not possible and one should take into account the possibility that the applied quantum chemical model deviates more strongly from reality than expected.

\subsection{Tetramantane}
\begin{figure}[t]\begin{center}
\epsscale{1.15} \plotone{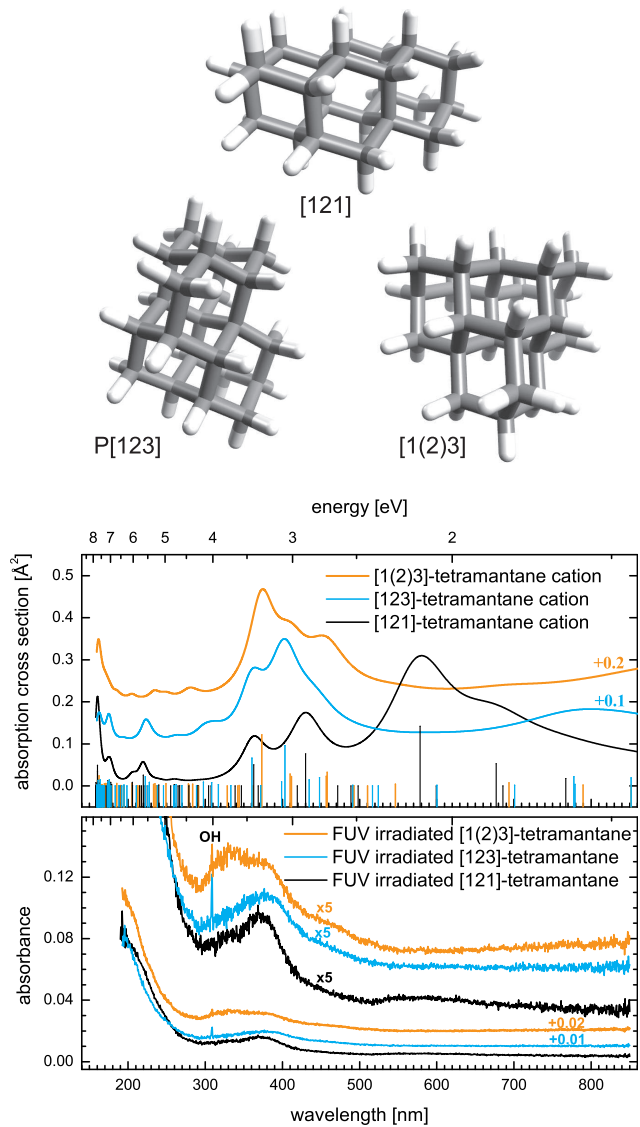} \caption{Isomers of tetramantane, calculated spectra of their cations, and measured spectra of their photoproducts isolated in solid Ne (6.8 K).} \label{fig8}
\end{center}\end{figure}

In contrast to the smaller diamondoids, there are already three different isomers of neutral tetramantane C$_{22}$H$_{28}$, while one of them has actually two enantiomers (P and M [123]-tetramantane). Their structures, the calculated spectra of their cations, as well as the measured spectra of their photoproducts, are displayed in Figure \ref{fig8}. We did not calculate the structures and spectra of the tetramantyl cations because of the increasing number of isomers and therefore escalating computational effort. Regarding the necessary energies for H abstraction, we do not expect large deviations from the smaller diamondoids. With almost no difference among the three species, the measured spectra are very similar to what has been measured for triamantane. Besides [121]-tetramantane, the broad feature extending roughly from 300 to 500 nm could be assigned to the open-shell cations as is obvious upon comparison with the calculations. The [121]-tetramantane cation, however, should have stronger bands at longer wavelengths suggesting that the measured spectrum is actually caused by the corresponding singly dehydrogenated cation.\footnote{On the other hand, it should be taken into account that the rise in the baseline beyond 510 nm could indicate a very broad and, compared to the calculated spectrum, rather weak band around 570 nm.} Whether the other two tetramantanes ([123] and [1(2)3]) lost a peripheral H atom upon ionization or not, cannot completely be clarified, as there are too many tetramantyl isomers and, as is the case with triamantane, a comparison with TD-DFT theory would not provide unambiguous insights.

\subsection{Complete $\sigma - \sigma^*$ Absorption Spectra}

\begin{figure*}[t]\begin{center}
\epsscale{1.15} \plotone{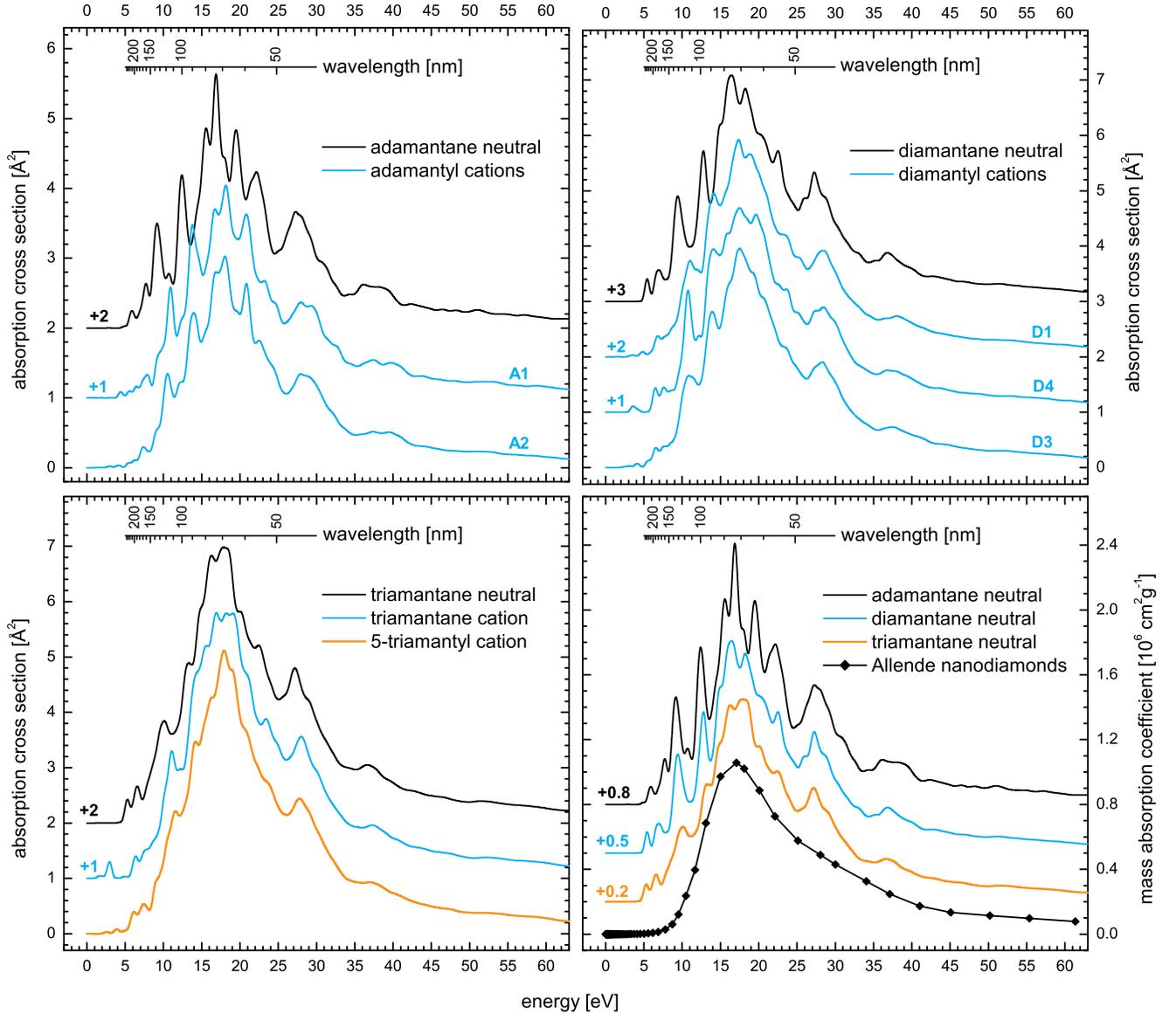} \caption{Complete calculated electronic $\sigma - \sigma^*$ absorption spectra of neutral and ionized small diamondoids. For comparison, the bottom right panel contains the IR to VUV spectrum of meteoritic nanodiamonds from the Allende meteorite ($\sim$ 2 nm) as derived from combined absorption and EELS (electron energy loss spectroscopy) measurements \citep{mutschke04}. The IR bands of the nanodiamonds are actually too weak compared to the electronic absorption to be seen in this scale.} \label{fig9}
\end{center}\end{figure*}

The electronic spectra of the neutral and cationic diamondoids resulting from (all possible) bound-bound ($\sigma - \sigma^*$) transitions, as calculated with the Octopus code, are displayed in Figure \ref{fig9}. These spectra may be used to model photophysical interactions of diamond-like molecules in the interstellar medium. For a brief discussion about their physical relevance refer to Section \ref{theory}. Regarding the energy range below 8.5 eV, more detailed, measured gas phase absorption spectra of neutral diamondoids can be found in the publication of \citet{landt09b}. The vibrational structure that can be seen in these spectra can hardly be predicted with current theoretical methods. The spectra presented here feature broader bands that are purely artificial. An additional energy-dependent broadening by convolving the spectra with Lorentzians of increasing bandwidths has been applied to account for an increased lifetime broadening which is expected at higher energies. Due to the high density of states above $\sim$ 10 eV, changing the bandwidths does not substantially alter the absolute cross section values. While the positions and shapes of resonances appearing in the spectra may be affected by uncertainties of the computational method the general trend of the absorption curves and the cross section values may be regarded as real (at least within the limitations discussed before). Comparing the spectra of the neutral and ionized molecules with each other, it is obvious that there are not many differences, especially for the transitions at higher energies, as the electronic structure of the C skeleton is equivalent. For all species, the high-energy absorption is dominated by a broad hump with a maximum between 15 and 20 eV. Also other features on the red and blue tails of this $\sigma - \sigma^*$ hump, like peaks at 11, 14, and 28.5 eV for the cations or 9, 12.5, and 27.5 eV for the neutrals are very much comparable. As already discussed in the previous sections, the absorption onset of the cations appears further to the red compared to their neutral precursors which is also evident from Figure \ref{fig9}. By increasing the molecular size, two effects are obvious: rising values for the absolute absorption cross section and a trend toward a less-structured absorption curve due to an increased density of states.

We want to point out an interesting aspect of these results. Even though laboratory experiments in the discussed energy range for these molecular species are lacking, experimental data on nanodiamonds, extracted and isolated from the Allende meteorite \citep{mutschke04}, display surprising resemblances (see Figure \ref{fig9}). These nanodiamonds possess an average size of less than 2 nm, corresponding to $\approx$ 500 C atoms, which is much bigger than the molecular diamonds presented here, the largest of which contains 22 C atoms. Their electronic absorption spectra solely consist of the broad $\sigma - \sigma^*$ band with maximum at 17.1 eV. Furthermore, a shoulder can be seen around 30 eV which may have its equivalent in a band at 28.5 eV in the diamondoid spectra. The peak mass absorption coefficient $\kappa$ for the meteoritic nanodiamonds was found to be $1.1 \times 10^6 \text{ cm}^2\text{ g}^{-1}$, very close to the calculated values of $\kappa = 1.2 - 1.3 \times 10^6 \text{ cm}^2 \text{ g}^{-1}$ for the molecular diamond. The main differences of the molecular compared to the nanoscopic material are the redshifted absorption onset and a more structured absorption curve, especially on the red wing of the collective $\sigma - \sigma^*$ hump.

\section{SUMMARY}
\label{summary}
We have investigated the electronic absorption properties of the smallest diamondoids, a possible molecular part of the interstellar carbonaceous dust with diamond-like structure. For adamantane and diamantane, our results confirm the formation of closed-shell singly dehydrogenated cations upon FUV irradiation which, similarly, was already found in previous studies by \citet{polfer04} and \citet{pirali10} in the infrared regime using an indirect ionization method. Further ionization of the adamantyl and diamantyl cations, even in strongly irradiated regions of space, may be hampered by the rather large second IP. For instance, DFT calculations (B3LYP / 6-311++G(2d,p)) imply  a 14.1 eV energy difference between the cation and dication of 1-adamantyl.  A clear identification of the created isomers in the matrix experiments, i.e., information about which H atom was removed from the edge of the molecule, through comparison with TD-DFT calculations is rather uncertain. Starting with triamantane, our results leave open the possibility that larger diamondoids are just ionized upon FUV irradiation, and that the absorbed photon energy is distributed over sufficient vibrational degrees of freedom to avoid H abstraction. Further spectroscopic investigations of FUV-processed diamondoids in the IR range would help to clarify this issue as molecular vibrations can be more accurately predicted by current theoretical models.

In our experiments, the dissociation of adamantane and diamantane was initiated via photoionization. Besides recombination reactions due to the close proximity of the molecules in the matrix, we do not expect that the interaction with the matrix atoms has a strong influence on the dissociation process itself and suggest that, if present as gas phase molecules, small diamondoids in interstellar space are subjected to H abstraction upon FUV irradiation. The measured spectra of the created ions isolated in solid Ne matrices display broad bands in the UV, shifting to longer wavelengths with increasing molecular size. The widths of these bands were found to be much larger than expected from typical broadening effects caused by the interaction with the rare gas atoms of the matrix. Therefore, we attribute this to an intrinsic effect of the molecules, i.e., a very short lifetime of the excited state which is not entirely caused by the interaction with the Ne matrix. This implies that, besides a small matrix-induced redshift, UV spectra of cationic diamondoids in the gas phase would feature bands similar in shape and width as in the matrix spectra. Unfortunately, the lack of narrow bands also hampers a possible detection in space via UV observations. The photoprocessed adamantane displays a broad absorption band at 223.5 nm which may be found at slightly shorter wavelengths in the gas phase. However, a possible contribution to the interstellar 217.5 nm UV bump can be rather excluded due to its weakness ($f \approx 0.09$) and several further absorption bands in close proximity at shorter wavelengths which are incompatible with the interstellar extinction curve. However, the specific structures of the dehydrogenated cations open other possibilities for detection. Unlike their neutral precursor molecules, such species possess strong permanent dipole moments due to the missing H atom and localized charge at the molecular periphery, permitting an identification by means of rotational spectroscopy. Potential targets for radio-based observations could be (the edges of) dense molecular clouds where small diamondoids can be expected, if the assignment of the 3.47 $\mu$m absorption band is valid, or the close proximity of objects with intense UV radiation fields, such as HD 97048 and Elias 1, where the 3.43 and 3.53 $\mu$m emission features were observed.

The main UV absorption feature at higher energies ($>$10 eV) of diamond-like material is the collective $\sigma - \sigma^*$ peak. Also the diamondoids and their cationic derivatives exhibit this feature and, interestingly, its position (18 eV) and cross section (max.  $1.2 - 1.3 \times 10^6 \text{ cm}^2 \text{ g}^{-1}$) are very similar to the extinction hump measured for the much larger meteoritic nanodiamonds \citep{mutschke04}. However, the molecular material additionally displays certain narrow features on the red tail of the $\sigma - \sigma^*$ hump \citep[see also ][]{landt09b}. These results may prove useful for the modeling of diamondoid IR emission caused by stochastic heating from UV photons and, thus, they can contribute to the discussion about the origin of the IR emission features of HD 97048 and Elias 1.\\

This work was supported by the Deutsche Forschungsgemeinschaft (DFG) and (in part) by the Department of Energy, Office of Basic Energy Sciences, Division of Materials Sciences and Engineering, under contract DE-AC02-76SF00515. M.S. thanks Dr. Harald Mutschke for providing the nanodiamond spectrum and for measuring the UV flux of the H$_2$ discharge lamp together with Kamel A. Khalil Gadallah.\\

\end{document}